\documentclass[%
 aip,
rsi,%
 amsmath,amssymb,
 reprint,%
]{revtex4-1}

\usepackage{graphicx}
\usepackage{dcolumn}
\usepackage{bm}
\usepackage{xcolor}

\usepackage[colorlinks,citecolor=magenta]{hyperref}

\graphicspath{{./Figures/}}

\begin{document}

\title{Hydrodynamic interactions dominate the structure of active swimmers' pair distribution functions}

\author{Fabian Jan Schwarzendahl}
\affiliation{Max Planck Institute for Dynamics and Self-Organization, Am Fa{\ss}berg 17, 37077 G\"ottingen, Germany}
\author{Marco G. Mazza}
\affiliation{Interdisciplinary Centre for Mathematical Modelling and Department of Mathematical Sciences, Loughborough University, Loughborough, Leicestershire LE11 3TU, United Kingdom}
\affiliation{Max Planck Institute for Dynamics and Self-Organization, Am Fa{\ss}berg 17, 37077 G\"ottingen, Germany}

\date{\today}

\begin{abstract}
Microswimmers often exhibit surprising patterns due to the nonequilibrium nature of their dynamics. Collectively, suspensions of microswimmers appear as a liquid whose properties set it apart from its passive counterpart. 
To understand the impact of hydrodynamic interactions on the basic statistical features of a microswimmer's liquid, we investigate its structure by means of the pair distribution function. We perform particle-based simulations of microswimmers that include steric effects, shape anisotropy, and hydrodynamic interactions. 
We find that hydrodynamic interactions considerably alter the orientation-dependent pair distribution function compared to purely excluded-volume models like active Brownian particles, and generally decrease 
the structure of the liquid. Depletion regions are dominant at lower filling fractions, while at larger filling fraction the microswimmer liquid develops a stronger first shell of neighbors in specific directions, while losing structure at larger distances. 
Our work is a first step towards a statistico-mechanical treatment of the structure of microswimmer suspensions. 
\end{abstract}

\maketitle


\section{Introduction}
\label{Sec:intro}
Collections of microorganisms such as bacteria or microalgae show a myriad of
self-organized structures and patterns that are also governed by physical processes stemming from their nonequilibrium nature and mutual interactions. 
Some examples are self-concentration~\cite{SchwarzendahlSM2018,dombrowskiPRL2004},
complex interaction with solid surfaces~\cite{OstapenkoPRL2018,laugaBiophysJ2006,berkePRL2008},
bacterial turbulence~\cite{wensinkPNAS2012}, 
swarming~\cite{copelandSM2009}, spontaneous formation of spiral vortices~\cite{wiolandPRL2013},
and directed motion~\cite{wiolandNJoP2016}.
Many of these systems collectively resemble the behavior and structure of a fluid
and therefore it has been found useful to model these collections of organisms 
as an active or living fluid \cite{KochARFM2011,MarchettiRMP2013}. 
Furthermore, the methods of statistical mechanics of classical fluids have 
been found to be powerful tools also in nonequilibrium situations, such as the analysis of velocity correlations and energy spectra for 
mesoscale turbulence of bacteria~\cite{wensinkPNAS2012}, and even a local equilibrium Maxwell--Boltzmann approximation for sedimenting active colloids~\cite{EnculescuPRL2011}.

The structure of an equilibrium fluid is typically analyzed in terms of the pair distribution function~\cite{Allen1987,Gray1984theory},
and is a cornerstone of liquid state theory~\cite{Hansen1990theory}. 
For active fluids, some work on the structural properties 
in terms of pair distribution functions has been put forward~\cite{BialkeEPL2013,HartelPRE2018,HoellJCP2018,WittkowskiNJP2017,RiedelScience2005,alarconJMolLiq2013,KoglerEPL2015,YangPRE2017,PessotMP2018,LocatelliPRE2015,TheersSM2018,deMacedoJOPCM2018,PessotMP2018}.
An important finding is the anisotropic form of the pair correlation function 
of active Brownian particles (ABP) presented in~\cite{BialkeEPL2013,HartelPRE2018,HoellJCP2018,WittkowskiNJP2017}, 
which arises from their self-propulsion. 

Typically, pair correlation functions are strongly influenced by the mutual interactions between particles,
which for microswimmers such as bacteria or algae moving in an aqueous milieu include hydrodynamic interactions. 
In this work we analyze the influence of the hydrodynamic interactions on the orientation-dependent pair distribution function of active swimmers, by considering a recent model for biological swimmers which also includes shape anisotropy and steric effects~\cite{SchwarzendahlSM2018}.
In order to disentangle the effects of activity, hydrodynamic and steric interactions, we compare the pair distribution functions for our model to the case of active Brownian particles.  
We find that hydrodynamic interactions introduce considerably more structure to the pair distribution function at low filling fractions and suppress structures at high filling fractions.

The rest of this work is organized as follows.
Section~\ref{Sec:model} introduces our model and the setup of the numerical simulations. 
In Sec.~\ref{Sec:pairdistrfunct} we discuss the orientation-dependent pair distribution function, and its angular average: the radial distribution function. 
Section~\ref{Sec:results} presents the distribution functions for both our swimmer's model and for ABP.  Finally, in Sec.~\ref{Sec:conclusion} we collect our conclusions.

\section{Models and parameters}
\label{Sec:model}

\subsection{Stroke-averaged biological microswimmer}
\label{Sec:StrokeModel}

\begin{figure}
        \centering
        \includegraphics[width=0.9\columnwidth]{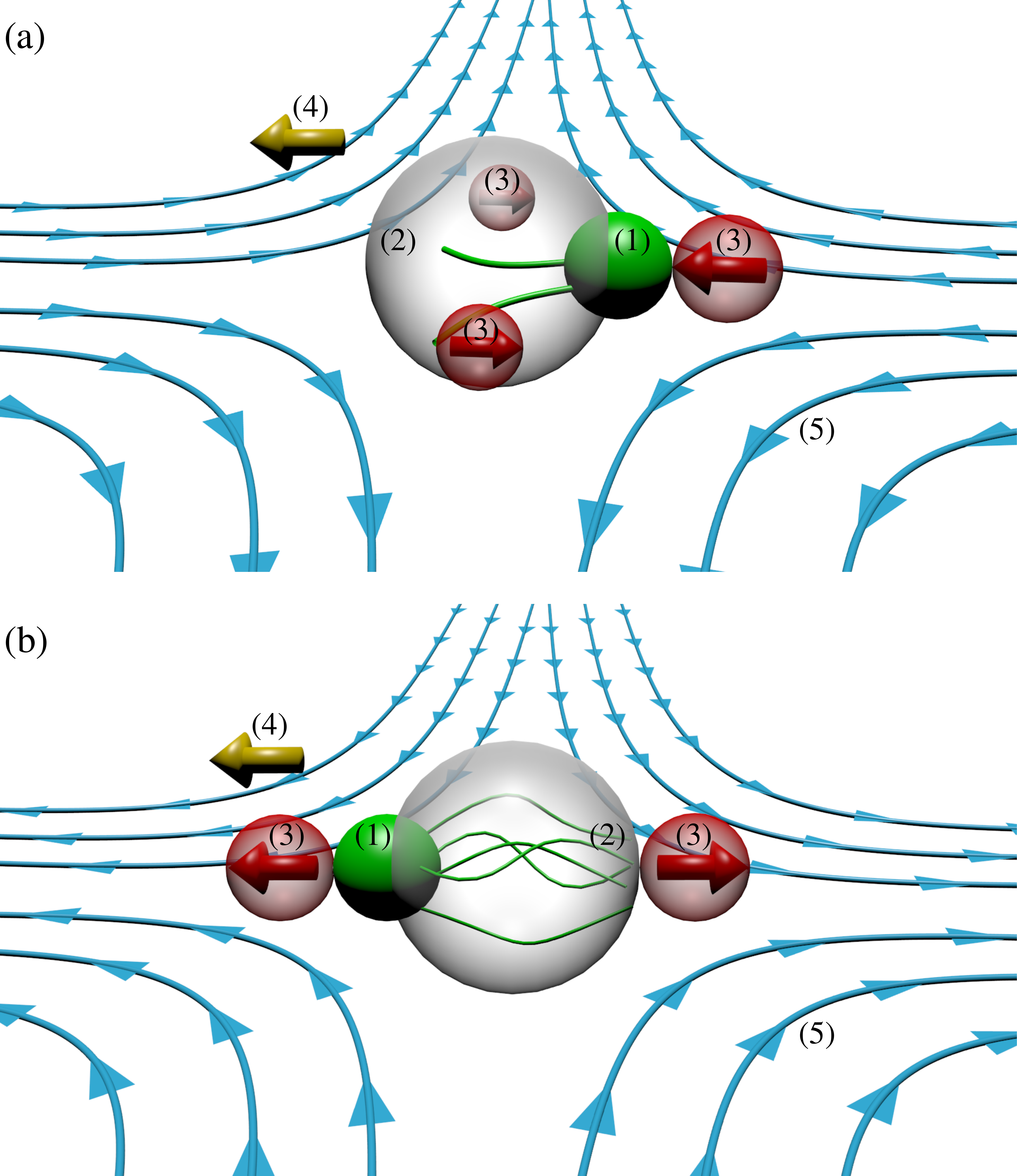}
        \caption{Schematic of our active swimmer model  (reprinted from \cite{SchwarzendahlSM2018}) 
        which shows the swimmers' composition, Stokeslets, and streamlines for a (a)  puller and (b) pusher-type swimmer.
        For both puller and pusher, the swimmer's body is represented by the small green sphere (1), and
        the stroke-average of the flagella is represented by the large transparent sphere (2).
        The regions in which the forces are applied are shown as red spheres with embedded arrows (3).
        The swimming direction is shown as the golden arrow (4). The hydrodynamic streamlines are sketched as the blue lines with arrows (5).}
        \label{fig:SwimmerModel}
\end{figure}

We employ the stroke-averaged biological microswimmer model presented in \cite{SchwarzendahlSM2018}, which consists of an asymmetric dumbbell.
Figure~\ref{fig:SwimmerModel}(a) shows a schematic representation of the puller-type model swimmer
mimicking a \textit{Chlamydomonas reinhardtii} cell, while
 Fig.~\ref{fig:SwimmerModel}(b) shows a schematic of the pusher-type model swimmer representing an \textit{Escherichia coli} cell. 
In both cases, the small green sphere, marked with (1), represents the swimmer body and the large sphere, marked with (2), represents the stroke-average region spanned by the flagellar motion.
The rigid motion of the dumbbells is governed by Newton's equations and implemented via  quaternion dynamics (for details see \cite{SchwarzendahlSM2018}).

As hydrodynamic interactions between microswimmers can play an important role in their collective behavior, we also include an explicit model for the surrounding fluid  by means of the multiparticle collision dynamics (MPCD) technique.
MPCD is a particle-based method that reproduces hydrodynamic modes up to the Navier--Stokes level of description \cite{gompperBookChap2009}.

Experimental measurements \cite{DrescherPNAS2011} have shown that the flow field of the pusher-type swimmer can be modeled by a force dipole,
whereas puller-type swimmers' flow fields are well described by a combination of three-Stokeslets \cite{DrescherPRL2010}.
We use regularized force regions, as shown by the red spheres with embedded arrows (3) in Fig.~\ref{fig:SwimmerModel}, to account for the hydrodynamic flow fields. 
Additionally, the dumbbells are coupled to the fluid by using a no-slip boundary condition on their surface (for more details see \cite{SchwarzendahlSM2018}).  

The temperature $T$ of the fluid, 
the size of an MPCD grid cell $a$, and the mass of an MPCD particle $m$, are the reference values for temperature, length, and mass, respectively. 
We perform simulations in a cubic domain of size $100a$ with periodic boundary conditions in all three directions;  
 each MPCD cell contains an average of $\langle N_\mathsf{C} \rangle = 20$ MPCD particles, 
giving a total of $2 \times 10^7$ MPCD particles in the system.
The effective radius of an individual swimmer is $\sigma \approx 5a$.
For further details on the numerical implementation the reader is referred to \cite{SchwarzendahlSM2018}.

\subsection{Active Brownian particles}
\label{Sec:ABP}

\begin{figure}
        \centering
        \includegraphics[width=0.4\columnwidth]{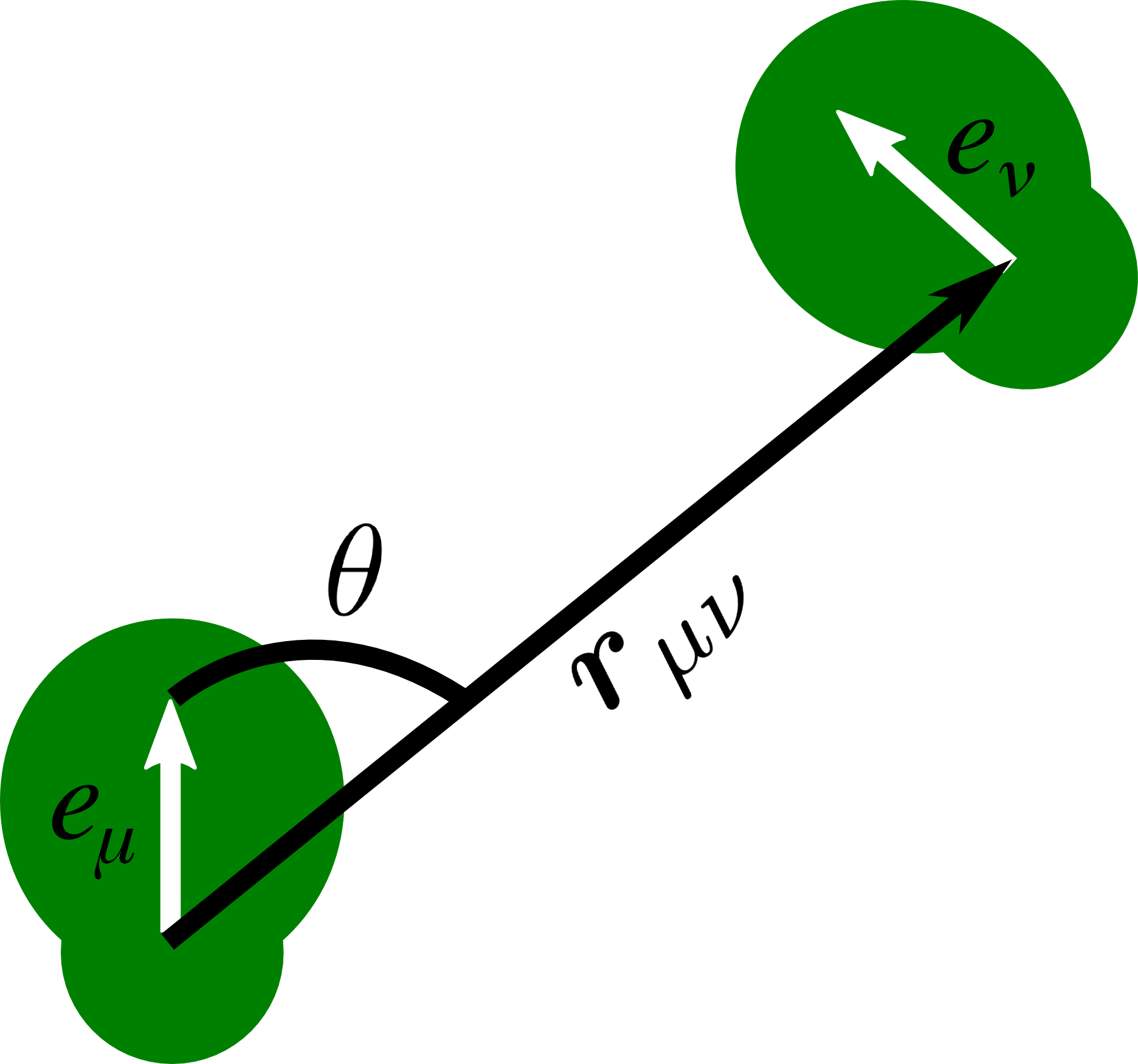}
                \caption{Sketch of two active swimmers with relative position $ \bm{r}_{\mu \nu} = \bm{r}_{\mu} - \bm{r}_{\nu}$,
                orientations $\bm{e}_{\mu}$, $\bm{e}_{\nu}$ and polar angle given by
                $\cos (\theta) = \bm{e}_{\mu} \cdot \bm{r}_{\mu \nu}/|\bm{r}_{\mu \nu}|$ with respect to $\bm{e}_{\mu}$.
                For the evaluation of the pair distribution function $g(r,\theta)$ the polar angle 
                $\theta$ and distance $r = |\bm{r}_{\mu \nu}|$ are
                computed for all distinct pairs of swimmers.}
        \label{fig:correlationexample}
\end{figure}

In addition to the our stroke-averaged biological microswimmer, we also carry out  simulations of ABP,
to disentangle the effects of activity, hydrodynamic and steric interactions. 
ABP are spherical particles with radius $\sigma$ that self-propel along their orientation $\bm{e}$ with a typical speed $v_0$ 
(see also \cite{WysockiEPL2014,FilyPRL2012,RednerPRL2013,BialkePRL2012}).
The translational motion of the particle's position $\bm{r}$ is governed by the equation
\begin{align}
        \frac{\text{d} \bm{r}}{ \text{d} t} = v_0 \bm{e} + \bm{F}/\gamma + \bm{\eta},
        \label{eq:BDtranslational}
\end{align}
where the force $\bm{F}$ between the particles is purely steric and is calculated from a Weeks--Chandler--Anderson potential \cite{WeeksJCP1971}.
Here, $\bm{\eta}$ is a Gaussian white noise with zero mean and $\langle \bm{\eta}(t) \bm{\eta}(t') \rangle = 2 D \bm{1} \delta(t-t') $, 
where $D$ is the translational diffusion coefficient.
The motion of the orientational degrees of freedom is determined by the equation
\begin{align}
        \frac{\text{d} \bm{e}}{ \text{d} t} = \bm{\zeta} \times \bm{e},
        \label{eq:BDorientational}
\end{align}
where rotational diffusion is included by the Gaussian white noise $\bm{\zeta}$ with $\langle \bm{\zeta}(t) \bm{\zeta}(t') \rangle = 2 D_r \bm{1} \delta(t-t') $ and
the rotational diffusion coefficient is given by $D_r= 3 D/(2\sigma)^2 $.
Also for the ABP model, we use a three dimensional cubic domain with periodic boundary conditions.

\subsection{Dimensionless parameters}
\label{Sec:DimNum}
For both our stroke-averaged biological microswimmer and ABP we simulate $N=300 -1350$ active swimmers, resulting in filling fractions from $\phi = 0.05$ to $\phi= 0.25$, at a fixed volume $(100a)^3$.
A dimensionless quantification of the relative importance of the self-propulsion speed to diffusive processes  is the P\'eclet number, which we define as $\mathcal{P}=  { v_{\mathrm{eff}} \sigma}/{D}$,
where $v_{\mathrm{eff}}$ and $\sigma$ are the effective speed and the effective size of the swimmer, respectively, 
and $D$ the diffusion constant of the swimmer (for ABP $v_{\mathrm{eff}}= v_0$.)
We present results at $\mathcal{P} \approx 1.9 \times 10^3$ and $\mathcal{P} \approx 6 \times 10^2$.
The Reynolds number of the flow around our swimmers (measuring the ratio of inertial to viscous forces) is given by 
$\mathcal{R}=\sigma v_{\mathrm{eff}}\rho/\eta$, where $\rho$ and $\eta$ are the MPCD fluid's density 
and viscosity, respectively.
The simulated Reynolds numbers are $\mathcal{R}=0.06$ for high P\'eclet number ($\mathcal{P} \approx 1.9 \times 10^3$)
and $\mathcal{R}=0.02$ at lower P\'eclet number ($\mathcal{P} \approx 6 \times 10^2$).

\section{Pair distribution function}\label{Sec:pairdistrfunct}

In general, the complete information of a liquid's structure is encapsulated in the angular
pair distribution function $g(\bm{r}_{\mu \nu},\bm{e}_{\mu},\bm{e}_{\nu})$ 
of two particles $\mu$ and $\nu$, which depends on their 
relative position $ \bm{r}_{\mu \nu} = \bm{r}_{\mu} - \bm{r}_{\nu}$
as well as on their orientations $\bm{e}_{\mu}$, $\bm{e}_{\nu}$ \cite{Gray1984theory,Allen1987}. 
However, in the general case the arguments of the pair distribution function are high dimensional 
and it has thus been proven useful to restrict oneself to a partial descriptions of the orientational 
ordering \cite{Allen1987,Gray1984theory}. 
Since our active swimmers self-propel into the direction of their orientation $\bm{e}_{\mu}$,
we are expecting non-trivial correlations mainly with respect to this axis.
Therefore, we compute correlation of a particles orientation with respect to the relative
position of all other particles, as illustrated in Fig.~\ref{fig:correlationexample}.

In practice, we use the following definition of our 
orientation-dependent pair distribution function: given two particles $\mu$ and $\nu$ with 
relative position $ \bm{r}_{\mu \nu} = \bm{r}_{\mu} - \bm{r}_{\nu}$
and orientations $\bm{e}_{\mu}$, $\bm{e}_{\nu}$, we compute the distance 
$r = |\bm{r}_{\mu \nu}|$ and the polar angle $\theta = \bm{e}_{\mu} \cdot \bm{r}_{\mu \nu}/|\bm{r}_{\mu \nu}|$ 
with respect to the orientation of particle $\mu$ (Fig.~\ref{fig:correlationexample}).
The average over the ensemble, polar angle, and orientation of the neighboring swimmer $\bm{e}_{\nu}$ is obtained from a histogram $h(r_b,\theta_b)$  for all $N^2$ pairs of swimmers, with radial bins $r_b$ and polar bins $\theta_b$.
The orientation-dependent pair distribution function is then given by 
\begin{align}
    g(r_b,\theta_b)= \frac{h(r_b,\theta_b)}{N^2 n(r_b,\theta_b)},
    \label{eq:radialdistribution}
\end{align}
where $n(r_b,\theta_b)$ is the normalization 
\begin{align}
    n(r_b,\theta_b) = &\frac{2}{3}\pi [ (r_b + \delta r)^3 - (r_b )^3 ] \nonumber \\
    & \times [ \cos(\theta_b) - \cos(\theta_b + \delta \theta)], 
\label{eq:radialnormalization}
\end{align}
where $\delta r$ is the bin size in the radial component and  $\delta \theta$ is the bin size in 
the polar component. 
It should be noted that $g(r,\theta)$ for a homogeneous configuration is equal to one.
Physically, $g(r,\theta)$ is related to the probability of finding a neighboring particle at distance $r$ and with a polar angle $\theta$ with respect to the reference particle and its orientation.


In addition to the orientation-dependent pair distribution function we calculate the radial distribution function \cite{Allen1987,Hansen1990theory}
given by
\begin{align}
    g(r) = \int_0^{\pi} g(r,\theta) \mathrm{d} \theta.
    \label{eq:rdfdef}
\end{align}
Furthermore, as an effective measure of the local structure around a microswimmer, we consider the coordination number \cite{Hansen1990theory}, which is the average number of nearest neighbours of a particle
\begin{align}
    C= \int_0^{R_\mathrm{min}} g(r) \mathrm{d}r,
    \label{eq:coordiantion}
\end{align}
where $R_\mathrm{min}$ is the position of the first minimum of $g(r)$.

\begin{figure*}
        \centering
        \includegraphics[width=2\columnwidth]{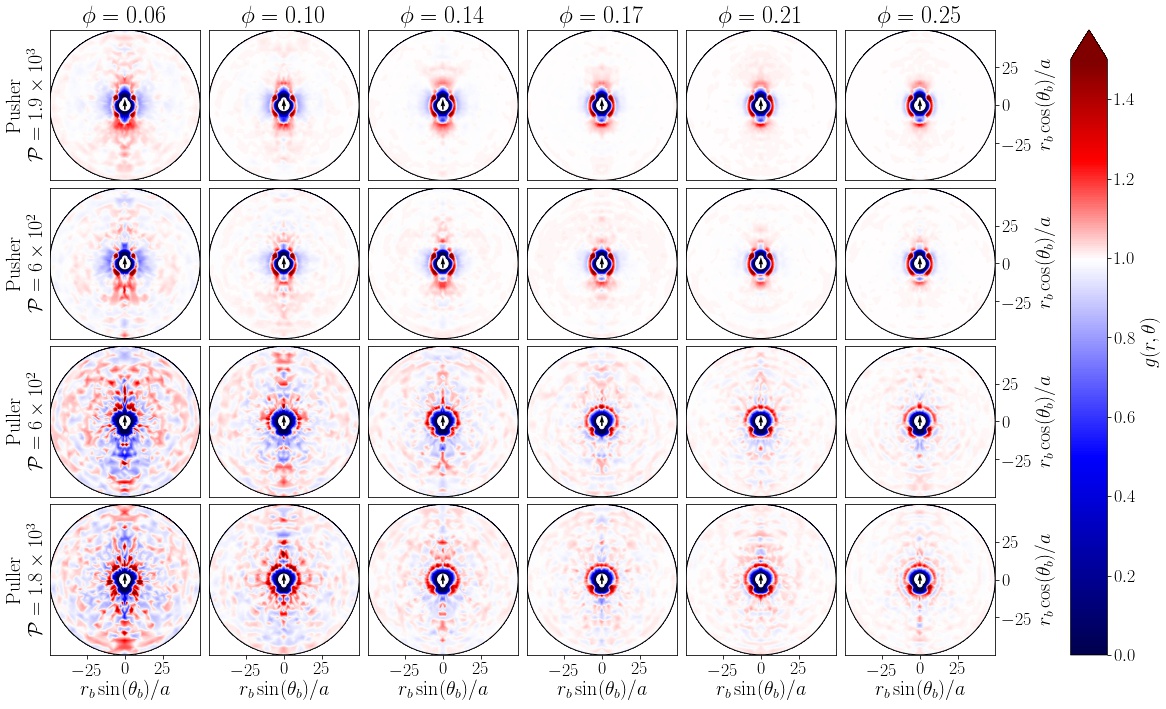}
                \caption{Orientation-dependent pair distribution function $g(r,\theta)$ for   pushers and pullers separated by a distance $r$ and with polar angle $\theta$.
                Each row shows a different P\'eclet number $\mathcal{P}$ and 
                each column a different filling fraction $\phi$. The color encodes the value of $g(r,\theta)$, where white indicates the bulk average $g(r,\theta)=1$. 
                The central, white dumbbells mark the position of the active swimmers and the black arrows the swimming direction. 
                }
        \label{fig:rdfmatrix1}
\end{figure*}
\begin{figure*}
        \centering
        \includegraphics[width=2\columnwidth]{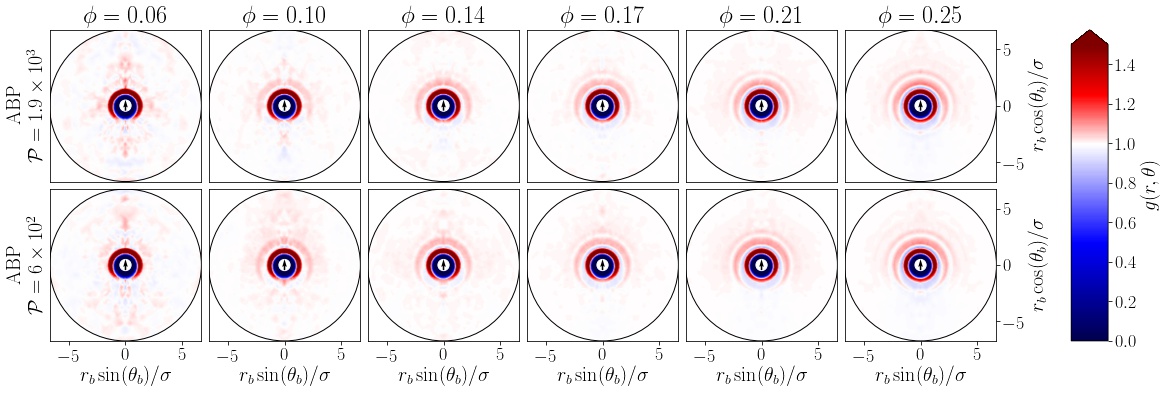}
                \caption{Orientation-dependent pair distribution function $g(r,\theta)$ 
                for ABP separated by a distance $r$ and with polar angle $\theta$. 
                Each row shows a different P\'eclet number $\mathcal{P}$ and 
                each column a different filling fraction $\phi$. The color encodes the value of $g(r,\theta)$, where white indicates the bulk average $g(r,\theta)=1$.
                The central, white circles mark the position of the ABP and back arrows the swimming direction. 
                }
        \label{fig:rdfmatrixBD}
\end{figure*}

\section{Results}\label{Sec:results}

Figure~\ref{fig:rdfmatrix1} shows a matrix chart of our calculations of [Eq.~\eqref{eq:radialdistribution}]. 
Different rows represent different P\'eclet numbers for both puller- and pusher-type swimmers, whereas  the  filling fraction changes with the column.
First of all, low filling fractions are characterized by a complex texture of peaks and minima of $g(r,\theta)$. These are stationary state structures, and not transient, as explicitly verified in our simulations.

Pushers have increased probability to find neighbors in the frontal and rear polar regions of their bodies, while around their equator neighboring swimmers tend to be depleted. As the P\'eclet number increases, the asymmetry between front and back poles increases, and the rear tail region grows more populated than the front. Pullers present the most complex texture at low filling fractions. A preference for an accumulation of probability in the front pole and in the rear region at intermediate latitudes is, however, discernible. Furthermore, at low filling fraction, $g(r,\theta)$
for both pullers and pushers shows a rather long-ranged structure, extending more than 5 times the radius of the individual swimmer.

As the filling fraction increases, $g(r,\theta)$ loses structure for both pushers and pullers at all P\'eclet numbers. 
For pushers, perpendicular to the swimming direction the pair distribution function first increases to $g(r,\theta)>1$
and then decreases to values $g(r,\theta)<1$. 
Furthermore, on the sides of the pushers one observes a band like structure for $\phi \gtrsim 0.10$ and a quadrupolar structure for $\phi = 0.06$.
The increased values of $g(r,\theta)$ in front and  back poles of  pushers 
 become more concentrated in smaller areas  as the filling fraction is increased.

For pullers, we observe that as $\phi$ increases,   $g(r,\theta)>1$ exhibits  a ring-like structure of increased probability close to the swimmers. Furthermore, we find two distinct regions at intermediate latitudes in the rear region of the swimmer, within the ring-like structure, that show very large values of $g(r,\theta)$.

To disentangle the effects of activity, steric and hydrodynamic interactions on the pair distribution function, 
we performed additional simulations of ABP~\cite{WysockiEPL2014,FilyPRL2012,RednerPRL2013,BialkePRL2012}, which only include steric interactions among swimmers.
Figure~\ref{fig:rdfmatrixBD} shows the orientation-dependent pair distribution function of ABP 
at different P\'eclet numbers and filling fractions. Even from a cursory inspection it is visible that the complex texture present at lower $\phi$ has nearly vanished. In its stead, a partial ring (interrupted in the lower pole) of pronounced peak of $g(r,\theta)$ emerges. 
A clear increase to $g(r,\theta)>1$ in front  
and a clear decrease to to $g(r,\theta)<1$ behind the active Brownian particle can be seen.
This asymmetry is an effect of the self-propulsion of the active Brownian particle and 
has also been studied in \cite{BialkeEPL2013,HartelPRE2018,HoellJCP2018,WittkowskiNJP2017}.

\begin{figure}
        \centering
        \includegraphics[width=1.0\columnwidth]{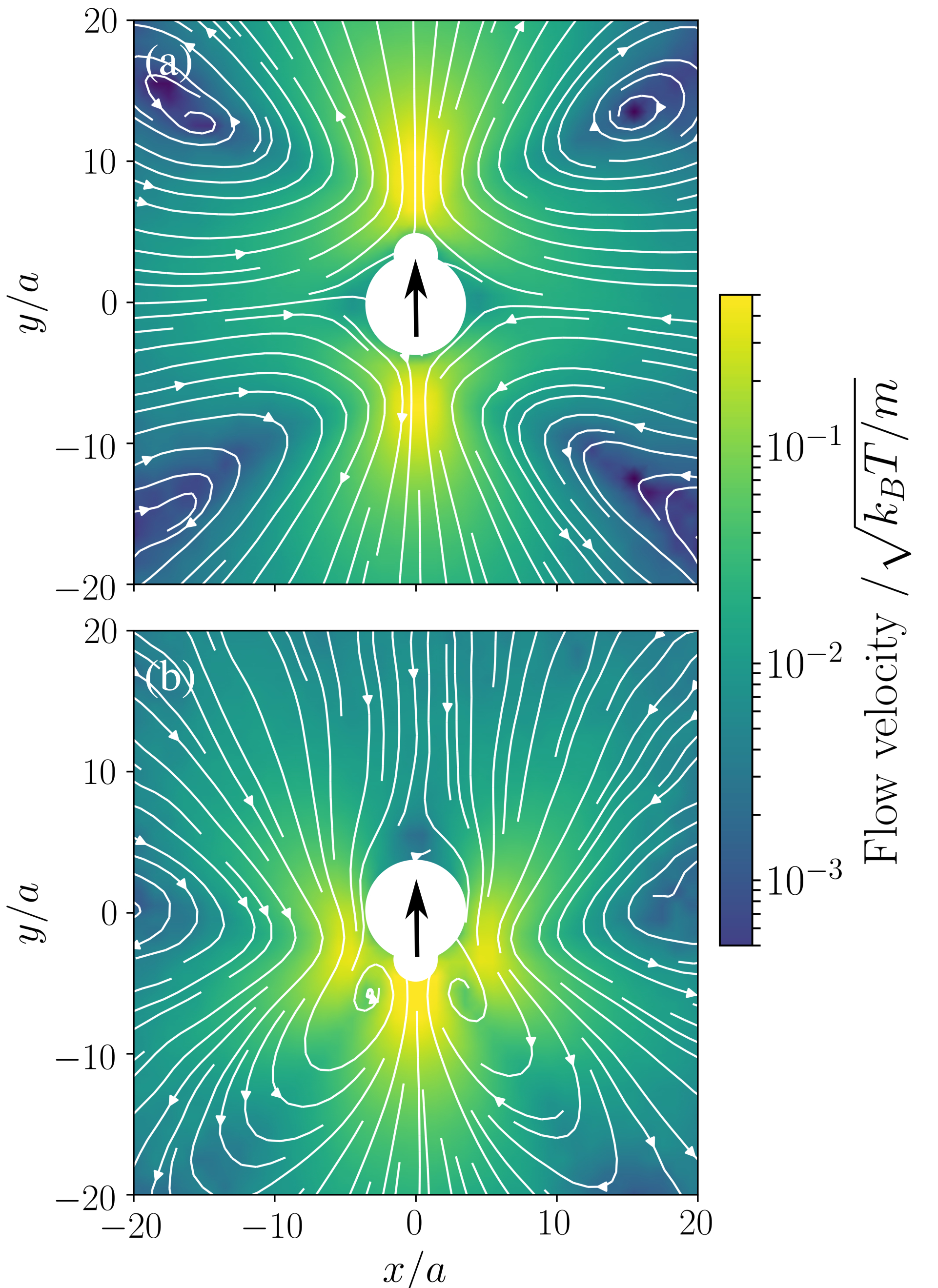}
                \caption{Flow field of an isolated (a) pusher and (b) puller. 
                Color code shows the flow velocity and white lines with arrows show the hydrodynamic streamlines. 
                The white dumbbells mark the position of the swimmers and the black arrow shows the swimming direction.}
        \label{fig:flow_fields}
\end{figure}

The comparison of active Brownian particles to our active swimmer model shows that the
hydrodynamic interactions have a strong impact on the pair distribution function.
The long-ranged structures that are observed in Fig.~\ref{fig:rdfmatrix1} at low filling 
fractions are not present for active Brownian particles in Fig.~\ref{fig:rdfmatrixBD}, 
and thus we conclude that these are a consequence of the hydrodynamic interactions between swimmers.

At high filling fractions, however, the opposite effect is observed: active Brownian particles show more
structure than our biological swimmer model, with a second and third order of rings of $g(r,\theta)>1$ in front and extending to the sides of the ABP. Again, this a consequence of hydrodynamic interactions.

The regions of increased probability [$g(r,\theta)>1$] around the equator of pushers, and in the rear region at intermediate latitudes for pullers strongly correlate with the structure of the respective flow fields. For pushers, the streamlines will bring a neighboring swimmer to the side of the one under consideration [see Fig.~\ref{fig:flow_fields}(a)]. For pullers, the stagnation point in the rear region at intermediate latitudes correlates well with the region of enhanced $g(r,\theta)$ [see Fig.~\ref{fig:flow_fields}(b)].
%
The strong influence from the specific flow field has also been recognized in \cite{PessotMP2018} in the context of a minimal microswimmer model.

\begin{figure}
        \centering
        \includegraphics[width=1.0\columnwidth]{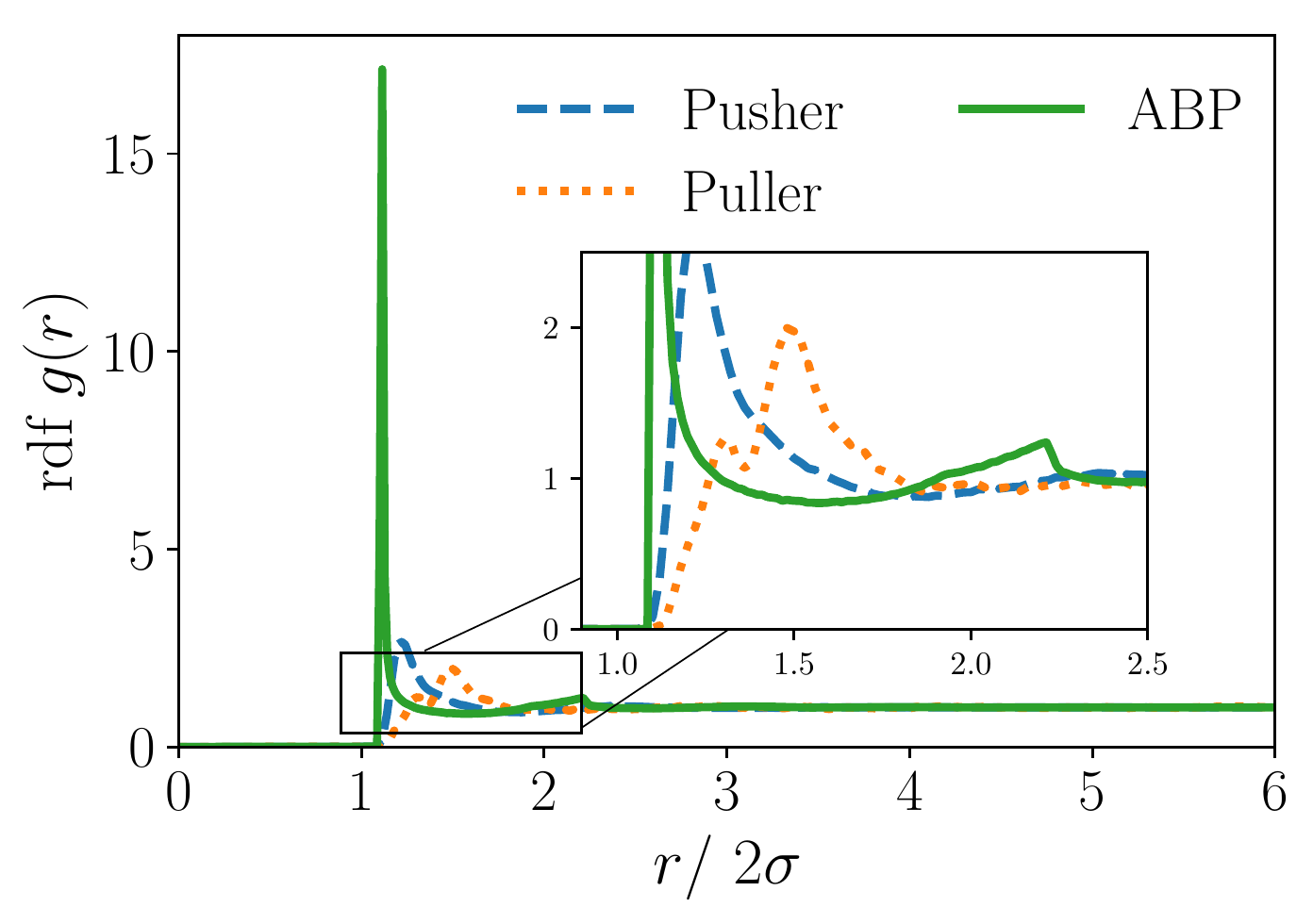}
                \caption{Radial distribution function for ABP, pushers, and pullers at filling fraction $\phi= 0.25$ and P\'eclet number $\mathcal{P}= 6 \times 10^2$. 
                        The inset shows a zoom in of the radial distribution function in the first peak region. The $r$-axis is scales by the typical particle diameter $2\sigma$ (for pullers and pushers we use $\sigma = 5a$)}
        \label{fig:rdf_example}
\end{figure}


Although the orientation-dependent pair distribution function already contains a coarse-grained level of description, complex structures are visible. It proves useful to observe an even more coarse-grained quantity $g(r)$, obtained from averaging over the polar angle $\theta$ and that will capture the average environment experienced by a swimmer over a long period of time. 
Figure~\ref{fig:rdf_example} shows the radial distribution function $g(r)$ for pushers, pullers, and ABP.
%
ABP exhibit a remarkably pronounced first peak and a much weaker second peak; a similar structure for ABP 
was also found by \cite{deMacedoJOPCM2018, faragePRE2015}.

The first shell of neighbors for both pushers and pullers is considerably weaker than the ABP's. This is a combination of the hydrodynamic interactions that quickly reorient particles, and the typically long reorientation times of ABP combined with steric effects~\cite{Matas-NavarroPRE2014}.
Even at this considerable level of coarse-graining, it is visible how the different flow fields of pushers and pullers generate different structures in $g(r)$. Pushers show only a single peak, whereas pullers have a weak first and a stronger second peak. 
Comparing pushers and pullers with ABP reveals that hydrodynamic interactions drastically reduce the height of the first peak but enhance its width. 

\begin{figure}
        \centering
        \includegraphics[width=1.0\columnwidth]{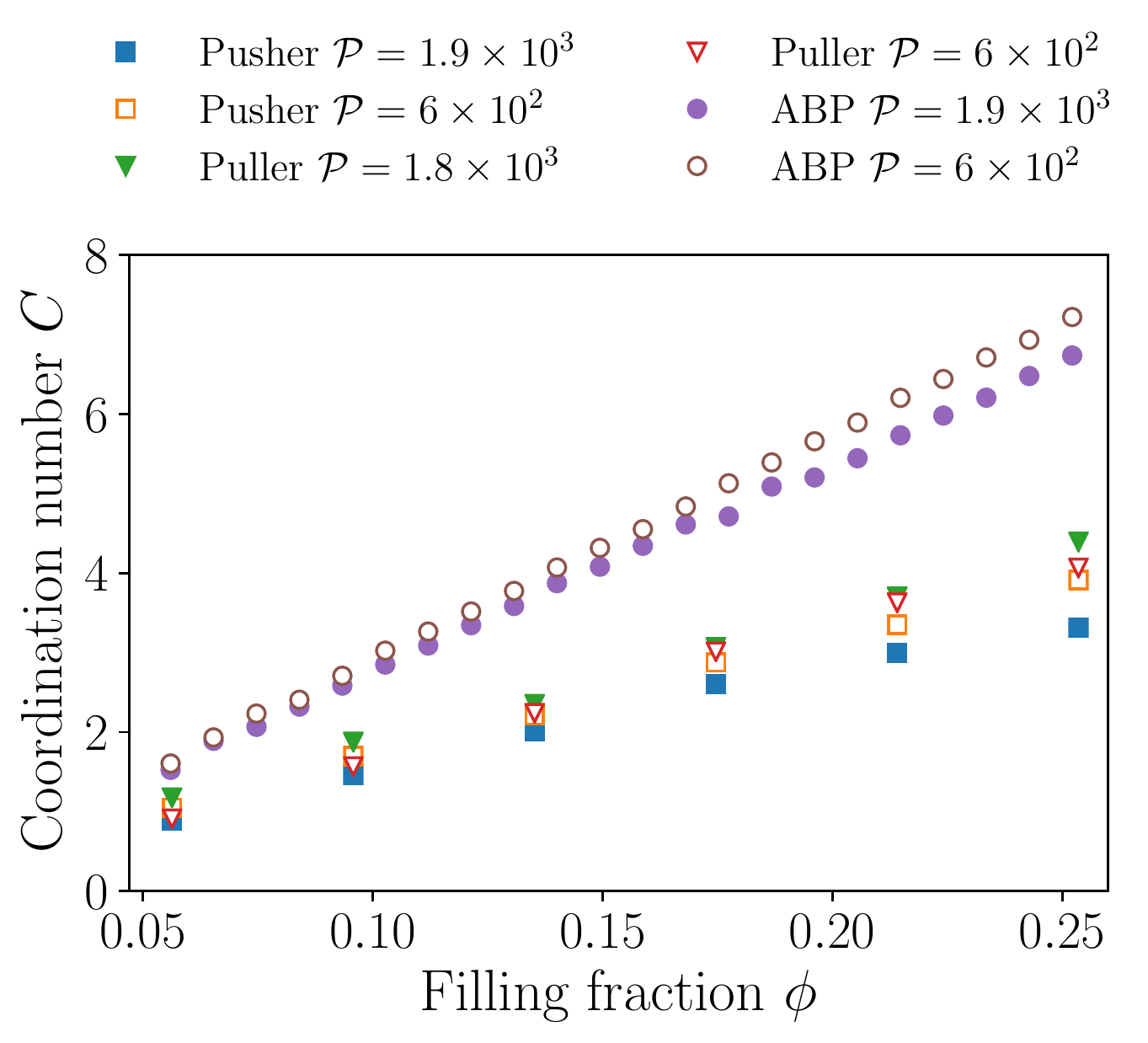}
                \caption{Dependence of the coordination number $C$ on the filling fraction $\phi$. 
                Pushers (squares), pullers (triangles) and ABP (circles) are shown for two different P\'eclet numbers.}
        \label{fig:first_coordination}
\end{figure}

To study the nearest-neighbor environment in more detail, we report the dependence of the  coordination number $C$ on the filling fraction $\phi$ in Fig.~\ref{fig:first_coordination}.
As $\phi$ increases, $C$ increases monotonically for pushers, pullers, and ABP at both P\'eclet numbers investigated.
However, for ABPs the coordination number is always larger than for pushers and pullers, 
confirming the result that the hydrodynamic interactions decrease the number of nearest neighbors.

\section{Conclusions}
\label{Sec:conclusion}
We have presented the orientation-dependent pair distribution function $g(r,\theta)$ for hydrodynamically interacting active 
swimmers at different P\'eclet numbers and a range of filling fractions. 
To disentangle the effects for activity, steric and hydrodynamic interactions we compared 
our results to simulations of active Brownian particles. 
We found that the hydrodynamic interactions introduce long-ranged structures into the 
orientation-dependent pair correlation function at low filling fractions. 
As the filling fraction increases, hydrodynamic interactions suppress the complex texture of $g(r,\theta)$. In fact, the influence of hydrodynamic interactions 
on the orientation-dependent pair correlation function is strong and the specific flow field introduces a non-trivial asymmetry in the $g(r,\theta)$  for both pushers and pullers. 

Investigation of the radial distribution function for pushers,  pullers, and ABP   
shows that hydrodynamic interactions strongly both suppress the height of the first peak and broaden it. 
As an effective measure of the average immediate neighborhood of a given microswimmer, the coordination number reveals  that  hydrodynamic interactions among
microswimmers suppress the strong first shell of neighbors. 

Our present work lays the foundations for a theoretical investigation of the complex interplay of hydrodynamics, steric forces, and active motion. 
An open ---and ambitious--- question is whether knowledge of the active liquid's structure is sufficient for the development of a nonequilibrium statistico-mechanical theory.

\section*{Conflicts of interest}
There are no conflicts of interest to declare. 

\section*{Acknowledgements}
We gratefully acknowledge support from the Deutsche Forschungsgemeinschaft (SFB 937, project A20).

\bibliography{JCP-ST}

\end{document}